\documentclass[aps,prl,twocolumn,groupedaddress,epsf]{revtex4}
\usepackage{epsf}
\usepackage{graphicx}

\begin{document}

\title{Two-Parameter Scaling of Microwave Rectification vs Microwave Power
at the Boundary between Two-Dimensional Electron Systems.}
\author{ N. Romero Kalmanovitz, I. Hoxha, Y. Jin, S. A. Vitkalov, M. P.
Sarachik}
\address{Physics Department, City College of the City
University of New York, New York, New York 10031}
\author{Ivan A. Larkin}
\address{International Center of Condensed Matter Physics, Bras\'{\i}lia, DF, 70904-970}
\author{T.~M.~Klapwijk}
\address{Delft University of Technology, Department of Applied Physics,
2628 CJ Delft, The Netherlands}

\date{\today }

\begin{abstract}
We report measurements of the rectification of microwave radiation ($0.7$-$%
20 $ GHz) at the boundary between two-dimensional electron systems separated
by a narrow gap on a silicon surface for different temperatures, electron
densities and microwave power. For frequencies above $4$ GHz and different
temperatures, the rectified voltage $V_{dc}$ as a function of microwave
power $P$ can be scaled onto a single universal curve $V_{dc}^{*}=f^*(P^{*})$%
. The scaled voltage is a linear function of power, $V_{dc}^{*} \propto P^{*}
$ for small power and proportional to $(P^*)^{1/2}$ at higher power. A
theory is proposed that attributes the rectification to the thermoelectric
response due to strong local overheating by the microwave radiation at the
boundary between two dissimilar 2D metals. Excellent agreement is obtained
between theory and experiment.
\end{abstract}

\maketitle


\section{Introduction}

\begin{figure}[tbp]
\vbox{\vspace{-1 in}} \hbox{\hspace{+0.1in}} \epsfxsize 3.4 in %
\epsfbox{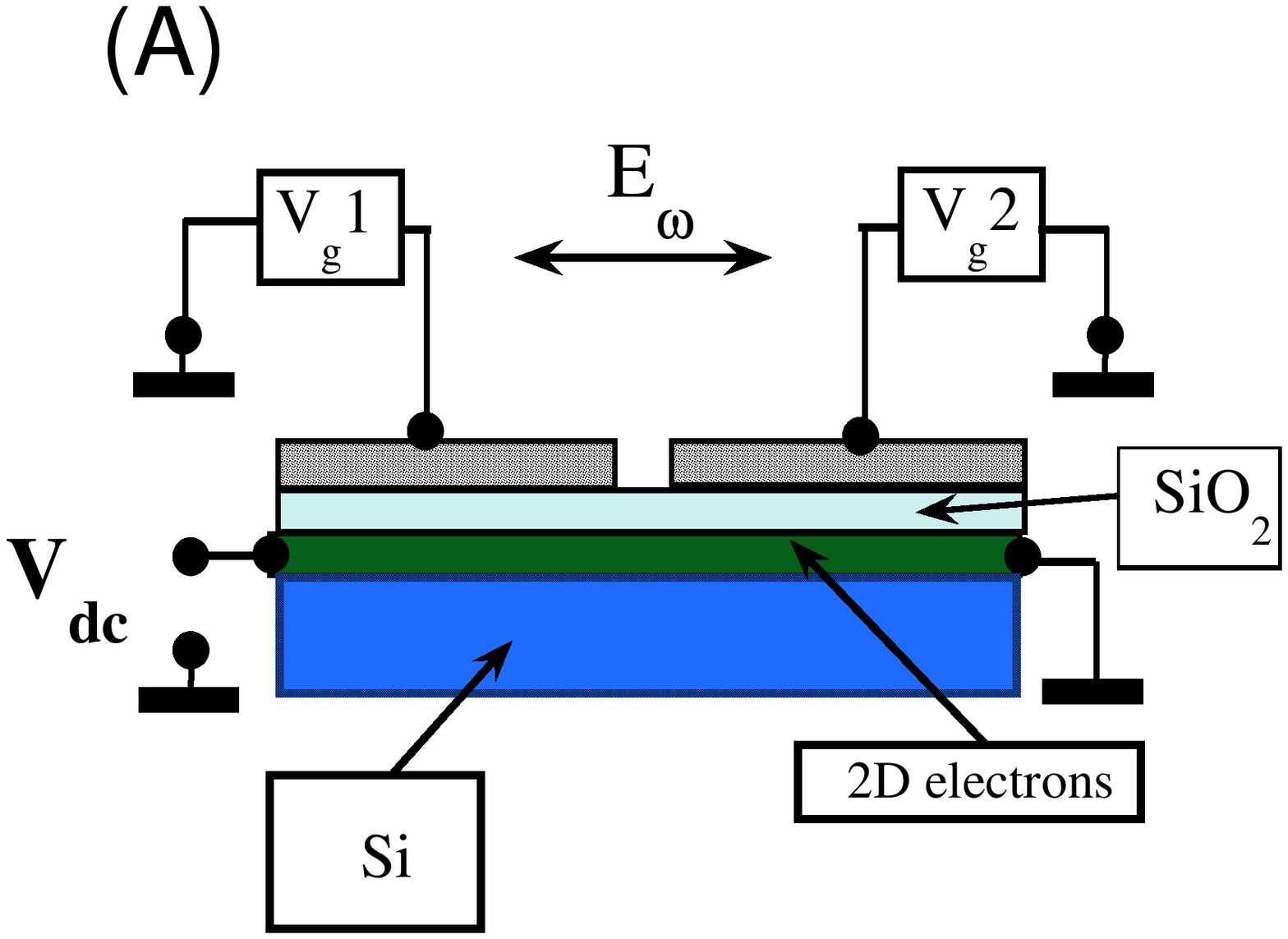} \vskip -3in \vbox{\vspace{0 in}} \hbox{\hspace{+0.1in}}
\epsfxsize 3.4 in \epsfbox{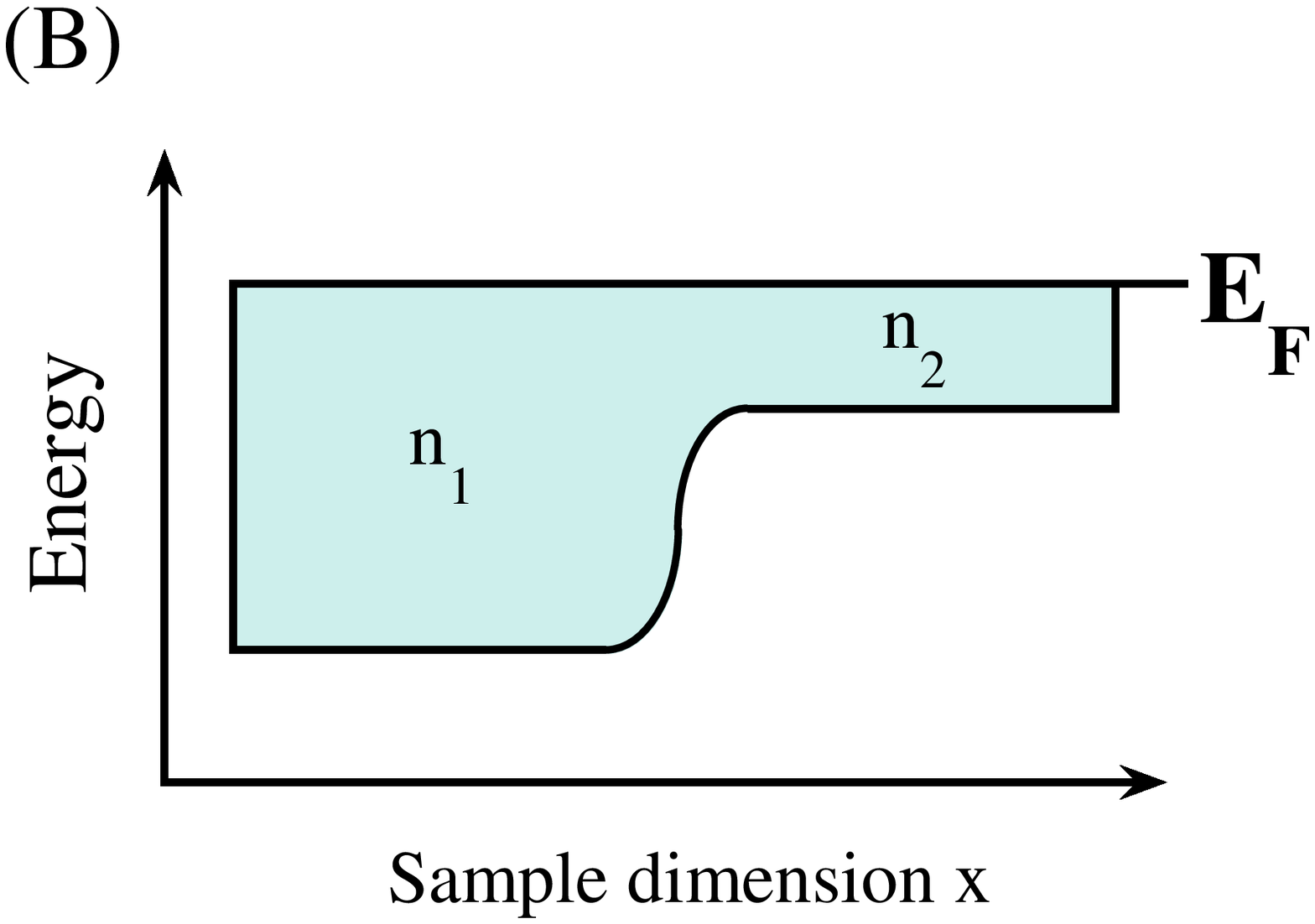} \vskip -1in
\caption{(A) Cross-section of the sample. Two different 2D metals are formed
under the two separate gates with voltages $V_g1$ and $V_g2$ applied as
shown. The rectified $dc$ voltage $V_{dc}$ is measured between the right and
left ends of the structure  \cite{contacts}. The microwave voltage is applied directly to the
gates and is localized near the slit between them (see Fig. \protect\ref{tx}%
); (B) the Fermi level and bottom of the conduction band are shown as a
function of position $x$ along the sample. Energies below the Fermi level
corresponds to occupied electron states of the two 2D metals with different
electron densities $n_1$ and $n_2$. The spatial variation of the electron
density is described by Eq. (1).}
\label{fig1_rect}
\end{figure}

The nonlinear behavior of low-dimensional electron systems has attracted a
great deal of attention for its fundamental interest as well as for
potentially important applications in nanoelectronics. In response to
microwave radiation and DC bias, strongly nonlinear transport \cite%
{zudov,engel,dorozh1,willett,stud1,bykov3,yang,bykov1,bykov2,zudov3,zudov4,
durst,anderson,andreev,shi,vavilov,dmitriev,auerbach,alicea,glazman,dmitriev2}
that gives rise to unusual states \cite{mani,zudov2,zdrs} has been reported
in two-dimensional systems of electrons in high magnetic field. There has
also been great interest in the nonlinear response of quantum ballistic
constrictions, where the effects of quantum interference, spatial dispersion
and electron-electron interactions play crucial roles \cite%
{dicarlo,wei,leturcq,zumbhl,lofgren,zhang,angers,brouwer,vavilovnl,sanchez,spivak,polianski,andreev2}%
.

In this paper we report a new type of nonlinearity of thermoelectric origin
in a two-dimensional system of electrons. We have investigated the rectified
(DC) voltage induced by microwave radiation applied locally to the boundary
between two-dimensional electron systems with different electron densities $%
n_1$ and $n_2$. A simple experimental geometry is used in which closely
spaced, electrically isolated gates give rise to a very large microwave
field localized at the narrow slit between the gates, thereby enhancing the
nonlinear response in the immediate vicinity of the gap. The gates are used
to vary the electron densities separately and independently, providing a
convenient and effective tool to control the strength of the nonlinearity.
The rectified voltage $V_{dc}$ is found to be an odd function of the
difference $\Delta n=n_2-n_1$ between the electron densities $n_1$ and $n_2$
of the two systems. Using two scaling parameters, all the data above $4$ GHz
taken at different temperatures can be collapsed onto a single universal
curve.

Excellent quantitative agreement is obtained with a theory that considers
the local overheating of the electrons by microwaves near the narrow
boundary between the 2D systems, which gives rise to a voltage through the
thermoelectric effect between two dissimilar two-dimensional metals. A fit
of the experimental data to this theory yields an electron-phonon coupling
constant that is in reasonable agreement with the coupling constant obtained
by a recent theory \cite{sergeev} as well as other experiments \cite{pudalov}%
.

This experimental protocol provides an effective method for studying the
thermoelectric properties of low dimensional systems. An important advantage
relative to other approaches \cite{pudalov} is that the electron system is
heated directly by the microwaves with negligible heating of the phonon
system, thereby reducing the contribution of phonon drag to the
thermoelectricity.

This paper is a continuation and expansion of research reported earlier \cite%
{hoxha,ivan1}. Although we attribute our observations to a thermoelectric
effect, we use the term ``microwave rectification'' for historical reasons.
The paper is organized as follows: the experimental set-up and measurements
are described in the next section; the following section summarizes the
experimental results; we then present a theory based on the thermoelectric
effect; this is followed by a discussion of other possible sources for the
observed rectification and a detailed comparison of data with the proposed
theory; the paper ends with an overall summary of our findings.

\section{Experimental Procedure}

\begin{figure}[tbp]
\vbox{\vspace{0 in}} \hbox{\hspace{+0.1in}} \epsfxsize 3.4 in \vskip -0.5in %
\epsfbox{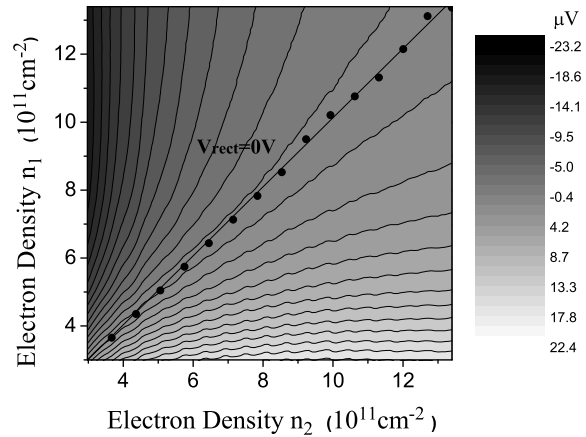} \vskip 0.5in
\caption{``Topological map" of the rectified voltage as a function of the
electron densities $n_{1}$ and $n_{2}$. The rectified voltage is indicated
by the shading, ranging from dark to light as the voltage varies from
negative to positive. Measurements were taken at temperature $4.2$ K,
frequency $20.2$ GHz, and microwave power input of $26$ dB (specified
relative to $1$ mW for $0$ dB). The rectification is an odd function of the
difference between the two densities $n_{1}$ and $n_{2}$ (see Eq. \protect
\ref{eqn1}). The black dots denote the values of $n_{1}$ and $n_{2}$ for
which the rectified signal changes polarity (goes through zero).}
\label{V_rect}
\end{figure}

The high-mobility Si-MOSFETs ($\mu =2$ m$^{2}$/(Vs) at $T=4$ K, $n=5\times
10^{15}$ m$^{-2}$) used in these studies are equipped with several metallic
gates that can be separately controlled. A narrow split obtained by reactive
ion etching separates the different gates. Each 2D electron system is formed
below a rectangular $50\times 240 \mu $m$^{2}$ gate by the application of a
positive voltage (see Fig. 1A). The typical slit width of $50-70$ nm is less
than the thickness of the Si oxide insulating layer ($152$ nm), providing a
smooth variation of the electron density between the two electron systems
formed below the gates, as shown in Fig. \ref{fig1_rect}(B). For a slit
width, $w$, that is much smaller than the distance $d$ between the gates and
the 2DEG, the profile of electron density is given by \cite{davies1}:

\begin{equation}
n(x)=\frac{n_{1}+n_{2}}{2}+\frac{n_{1}-n_{2}}{2}\tanh (\frac{\pi x}{d}),
\label{eqn0}
\end{equation}%
where $x$ is the distance from the center of the slit. For the actual
parameters of our samples, the exact solution differs from Eq. (\ref{eqn0})
by less than 3\%. The six different pairs of 2D electron systems studied
displayed similar behavior.

Measurements were taken at frequencies from $0.7$ GHz to $20$ GHz in a
vacuum chamber of a He-3 cryostat. The microwave radiation was guided by a
semi-rigid coaxial line terminated by a loop. Two wires, anchored to a
temperature controlled cold finger, were inductively coupled to the loop
without touching it. The wires were connected directly to the two adjacent
gates. The sample, with a calibrated RuO$_2$ thermometer attached, was
thermally connected to the same cold finger. The temperature of the
electrons was monitored using the amplitude of Shubnikov-de Haas (SdH)
oscillations. Without microwave power input, the electron temperature
followed the temperature of the cold finger down to the lowest temperature $%
T=0.27$ K. The sample was thus well isolated from heat input deriving from
the coaxial line and the RF filtered DC electrical leads.

Rather than the resistance, the dominant contribution to the output
impedance $Z_{out}$ of the circuit at the end of the microwave line was
provided by the (substantial) capacitive coupling between the two closely
placed wires. Estimates indicated that variations of the resistivity of the
2D electrons with temperature and gate voltage have a negligibly small
effect on $Z_{out}$. We therefore neglected the consequent small changes of
output voltage $V_{ac}$. In particular, we neglected the effect of
overheating the 2D electrons by the microwaves on output impedance and
consider the amplitude of the microwave voltage $V_{ac}$ applied to the
gates to be proportional to the square root of the calibrated microwave
power $P$ applied at the input of the coaxial line: 
\begin{equation}
V_{ac}=G\cdot P^{1/2},  \label{G}
\end{equation}%
where $G$ is a temperature independent coefficient.

\begin{figure}[tbp]
\vbox{\vspace{-0.2 in}} \hbox{\hspace{+0.1in}} \epsfxsize 3.4 in %
\epsfbox{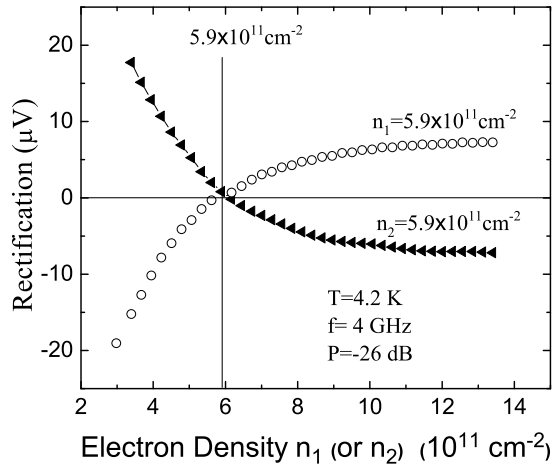}
\caption{The open circles denote the rectified signal as a function of the
electron density $n_{2}$ with electron density $n_{1}$ kept constant at $%
5.9\times 10^{11} cm^{-2}$. The triangles show the rectified signal versus
the electron density $n_{1}$ with the density $n_{2}$ fixed at the same
value $5.9\times 10^{11} cm^{-2}$. Measurements were taken at temperature $%
4.2$ K, frequency $4$ GHz, and microwave power $26$ dB (specified relative
to $1$ mW for $0$ dB). Note that if $n_{1}$ and $n_{2}$ are interchanged the
rectification changes sign and satisfies relation (\protect\ref{eqn1}).}
\label{Rect_n}
\end{figure}

The microwave-induced DC voltage was measured between two electrical
contacts placed on opposite sides of the sample at a distance $L=240 \mu$m
from the gap between the two electron systems (see Fig. 1 (A)). To avoid
thermoelectric effects related to the electrical contacts, the distance $L$
must be considerably longer the thermal relaxation length $L_T$, which is
estimated to be $L_T \sim$100 $\mu$m in Si-MOSFETs at temperatures of $\sim1$
K \cite{pudalov,prus,sergeev} (see Fig. \ref{tx}). In this paper we present
data obtained at temperatures above $2$ K, where the contact
thermoelectricity is negligibly small.

The same results for the rectification were obtained using continuous
microwave radiation and by modulating the microwave amplitude at frequency
typically 10Hz and using standard phase sensitive techniques. All the data
reported in this paper were obtained by the second method, as it provided
better detection of signals below $1 \mu V$.

\section{Experimental Results}

\subsection{Dependence of the rectification on electron density}

\begin{figure}[tbp]
\vbox{\vspace{0 in}} \hbox{\hspace{+0.1in}} \epsfxsize 3.4 in %
\epsfbox{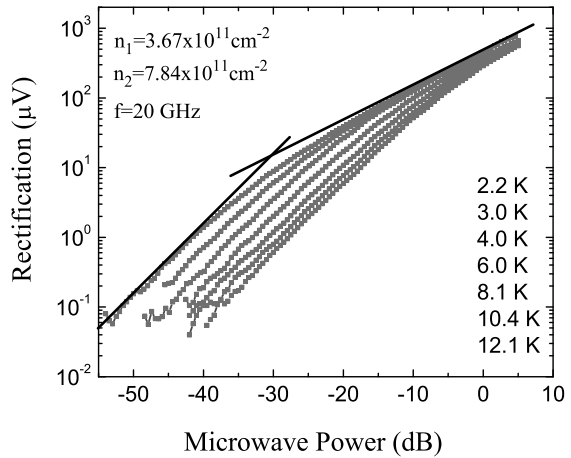}
\caption{Rectified voltage as a function of microwave power on a log-log
scale. The units of power are specified relative to $0$ dB for $1$ mW.
Curves are shown for seven different temperatures ranging from $2.2$ K (top
curve) to $12.1$ K (bottom curve). The straight lines represent linear $P$
and square root $P^{1/2}$ dependence of the rectification on microwave power 
$P$. The electron densities are $n_1=3.67 \times 10^{11}$ cm$^{-2}$, $n_2=7.84
\times 10^{11} $cm$^{-2}$. The microwave frequency is $20$ GHz.}
\label{Rect_P}
\end{figure}

\begin{figure}[tbp]
\vbox{\vspace{0 in}} \hbox{\hspace{+0.1in}} \epsfxsize 3.4 in %
\epsfbox{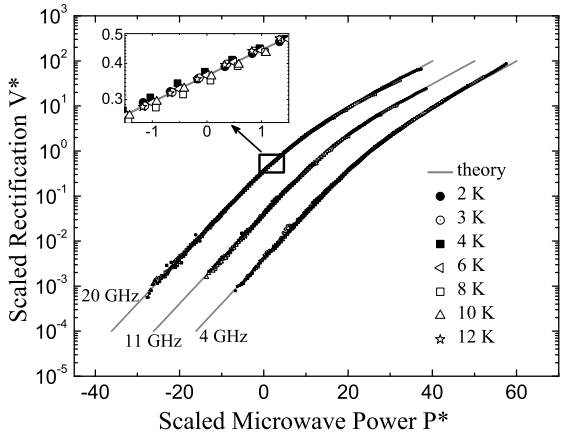}
\caption{ Normalized rectified voltage $V_{dc}^{*}$ versus normalized
microwave power $P^{*}$ for frequencies 20 GHz (the data of Fig. \protect\ref%
{Rect_P}), $11$ GHz and $4$ GHz at different temperatures. The data for
frequencies $11$ GHz and $4$ GHz are shifted horizontally by $10$ dB and $20$
dB respectively with respect to the curve at $20$ GHz for clarity. The solid
lines are theoretical curves given by Eq. \protect\ref{eqn_rect_norm} and
Eq. \protect\ref{y0_norm} (also shifted by $10$ dB and $20$ dB). The inset
is a magnification of a portion of the top curve to indicate the quality of
the scaling and/or typical deviations from the theory. A scaled curve is
obtained at each frequency for all temperatures, and an appropriate
horizontal shift brings the curves into alignment onto a single universal
curve for all frequencies shown.}
\label{scaling}
\end{figure}

Figure \ref{V_rect} shows the dependence of the rectified signal on the
electron densities of the two adjacent 2D metals. The axes denote the
electron densities $n_1$ and $n_2$, and the shading reflects the amplitude
of the rectified signal, with dark (light) shading denoting negative
(positive) values. Each horizontal scan was obtained for a fixed density $n_1
$ (the electrons in the left-hand region of Fig. 1(A)), while the electron
density $n_2$ in the adjacent (right-hand) region is varied. Shown by the
black dots in Fig. 2, the DC voltage changes sign when the electron
densities of the two 2D metals are nearly the same.

We note that similar results were obtained in our previous experiments \cite%
{hoxha}, where microwave radiation was applied by a very different method
using two parallel wires placed far from the samples. This indicates that
the results are robust and do not depend on details of the distribution of
electromagnetic fields in the vicinity of the sample and that the
dissimilarity between the two metals rather than the microwave field
distribution is responsible for the effects observed. Moreover, the absence
of rectification when $n_1=n_2$ is strong experimental indication that bulk
rectification inside the 2D metals, associated with microwave modulation of
the electron density and/or mobility, is a minor contribution to the
observed signal. This is discussed further in subsequent sections of the
paper.

Figure \ref{Rect_n} provides a clear demonstration that the rectified signal
is an odd function of the difference $\Delta n=(n_{2}-n_{1})$ between the
electron densities of the two systems. Here curve (a) shows the
rectification when the electron density $n_{1}$ is fixed at $5.9\times
10^{11}cm^{-2}$ while the electron density $n_{2}$ is varied; curve (b) is
for fixed density $n_{2}=5.9\times 10^{11}cm^{-2}$ and variable density $%
n_{1}$. Almost perfect antisymmetry is found with respect to the horizontal
axis, showing that the rectification is an odd function of the difference
between the two electron densities:

\begin{equation}
V_{dc}(n_1-n_2)=-V_{dc}(n_2-n_1),  \label{eqn1}
\end{equation}

\begin{figure}[tbp]
\vbox{\vspace{0 in}} \hbox{\hspace{+0.1in}} \epsfxsize 3.4 in %
\epsfbox{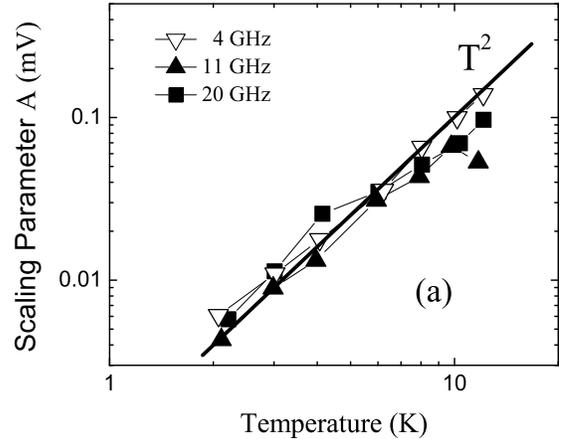} \vbox{\vspace{0 in}} \hbox{\hspace{+0.1in}} \epsfxsize %
3.4 in \epsfbox{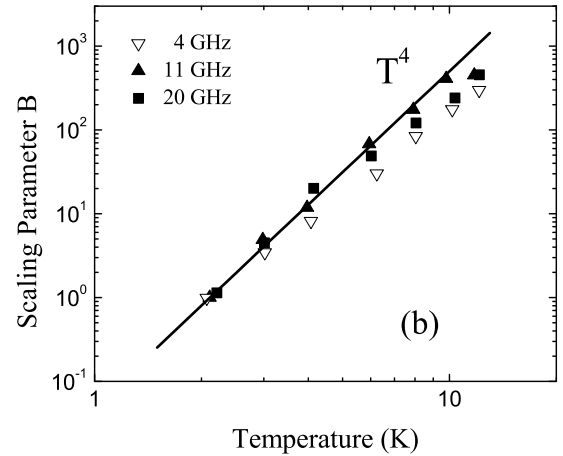}
\caption{ (A) Dependence of scaling parameter $A$ on temperature at
different frequencies, as labeled. The units of $A$ are chosen to conform to
the theoretical prediction of Eq. \protect\ref{eqn_rect}. The solid straight
line is theoretical dependence corresponding to Eq. \protect\ref{A}; (B)
Dependence of scaling parameter $B$ on temperature at different frequencies
as labeled. Solid straight line is theoretical dependence corresponding to
Eq. \protect\ref{B} with the parameter $R \propto T_L^6$.}
\label{Tdep}
\end{figure}

\subsection{Dependence of the rectification on microwave power at different
temperatures}

\begin{figure}[tbp]
\vbox{\vspace{0 in}} \hbox{\hspace{+0.1in}} \epsfxsize 3.4 in %
\epsfbox{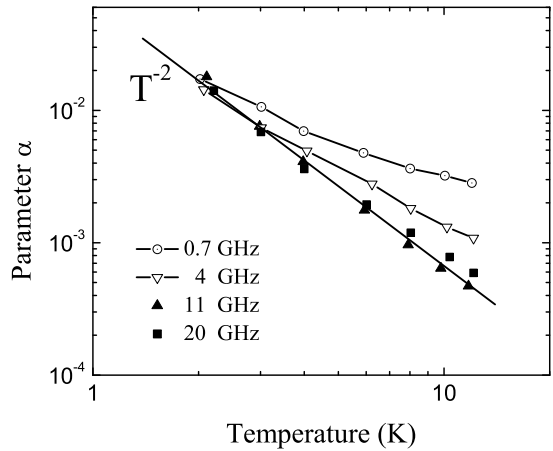}
\caption{ Temperature dependence of the proportionality constant, $\protect%
\alpha$, in the low microwave power regime where $V=\protect\alpha \times P$%
. For $11$ GHz and $20$ GHz, $\protect\alpha \propto T^{-2}$; the straight
line corresponds to the theoretical prediction (Eq. \protect\ref{eqn_rect}
and Eq. \protect\ref{y0}) at high frequencies. The electron densities are $%
n_1=3.67 \times 10^{11}$ cm$^{-2}$ and $n_2=7.84 \times 10^{11}$ cm$^{-2}$. }
\label{alpha}
\end{figure}

Figure \ref{Rect_P} shows the rectified voltage as a function of microwave
power on a log-log scale for seven different temperatures ranging from $2.2$
K to $12$ K at microwave frequency 20 GHz \cite{bfield}. At all
temperatures, the rectified signal for low power input is proportional to
the microwave power (the square of the microwave electric field), $
V_{dc}\propto P\propto E_{\omega }^{2}$. We will refer to this as the weak,
or perturbative, nonlinear regime. Strongly nonlinear behavior is observed
at higher levels of microwave excitation; here the rectified signal is
proportional to the square root of the power, $V_{dc} \propto P^{1/2}
\propto E_\omega$. The two straight lines drawn in Fig. \ref{Rect_P}
represent the two limits, namely, the perturbative ($V_{dc} \propto P$) and
the strongly nonlinear ($V_{dc} \propto P^{1/2}$) regimes. The crossover
between the two regimes depends on temperature, with the crossover occurring
at higher microwave power as the temperature is increased.

By applying appropriate multiplicative scale factors (corresponding to
translations on a log-log plot) to each of the curves of Fig. \ref{Rect_P},
one can collapse all the data for rectified voltage versus power at
frequency $20$ GHz onto the single universal curve:\ 

\begin{equation}
V_{dc}^{*}=f^*(P^{*}), ,  \label{eqn_propto}
\end{equation}
as shown in Fig. \ref{scaling}. Also shown in Fig. \ref{scaling} are
similarly scaled curves for frequencies $11$ GHz and $4$ GHz. Departures
from scaling that occur below $4$ GHz will be discussed later in this paper.

The scaled values of the rectified voltage $V_{dc}^{\ast }=V/A(T)$ and power 
$P^{\ast }=P/B(T)$, require the two scaling parameters $A(T)$ and $B(T)$
shown in Fig. \ref{Tdep} as a function of temperature for different
frequencies. For temperatures below $6$ K the parameters $A(T)\propto T^{2}$
and $B(T)\propto T^{4}$, with deviations toward a weaker dependence at
higher temperatures. There is no clear dependence on frequency. At a lower
frequency (0.7GHz) the scaling breaks down for high power input, with
substantial deviations from $V_{dc}\propto P^{1/2}$ behavior (not shown).

In the low power regime the rectification is found to be proportional to
microwave power at all frequencies. In this weakly nonlinear regime, the
rectification can be written as: 
\begin{equation}
V_{dc}=\alpha(T)\times P,  \label{lin_dep}
\end{equation}
where $\alpha(T)$ depends on the temperature. Figure \ref{alpha} shows the
constant of proportionality $\alpha$ plotted as a function of temperature
for different frequencies.

The scaling behavior indicates (see Fig. \ref{scaling}) that $%
V_{dc}^{*}=\gamma \cdot P^{*}$ in the weak nonlinear regime with a constant,
temperature-independent $\gamma$. Since the scaling parameters $A$ and $B$
are given by $V_{dc}^{*}=V/ A(T)$ and $P^{*}=P/B(T)$ it follows that the
coefficient $\alpha=\gamma \cdot \frac{A(T)}{B(T)}$. The solid line shows
the behavior expected from the theory in the scaling regime.

\section{Theory}

In this section we present a quantitative theory for the microwave
rectification in the two 2D electron systems studied in the previous
section. In this theory the DC voltage results from the thermoelectric
effect induced by local microwave overheating of the area near the boundary
between two dissimilar 2D electron metals. First we present a system of
electrodynamic equations, which allows us to find the distribution of
microwave electric potential across the sample. We will show that at high
frequency the microwave field is localized near the narrow gap between the
metals. Then we solve the thermoconductivity equation assuming fast
thermalisation of hot 2D electrons and find the temperature profile inside
the electron system. Deviations from the isotropic Fermi distribution should
not significantly affect the spatial relaxation and, therefore, will not be
discussed in this paper. The thermoelectric voltage $V_{dc}$ is found by
direct integration of the thermoelectric field across the sample.

In the derivation of the microwave ($ac$) current distribution, we take into
account (i) that the microwave wavelength ($0.7-30$ cm) is much larger than
the device size ($\sim 0.05$ cm) and, therefore, we can omit the term $%
\partial B/\partial t$ in Maxwell's equations; and (ii) that the scale at
which the electric potential varies $l_{ac}$ is much larger than the
effective distance of the 2D conducting layer from the gates $d$. Together
with the law of electric charge conservation, the full set of equations for
the time-dependent current distribution $j(x,t)$, density $\delta n(x,t)$\/,
and the electric potential at the 2D conducting plane reads: 
\begin{eqnarray}
j(x,t) &=&\sigma (x)\nabla \phi (x,t).  \label{field} \\
\partial _{t}[e\delta n(x,t)] &=&-\nabla j(x)  \label{field2} \\
\phi (x,t) &=&\phi _{0}(x,t)+\frac{ed}{\epsilon \epsilon _{0}}\text{\/}%
\delta n(x,t)\text{\/},  \label{field3}
\end{eqnarray}%
where $\sigma (x)$ is the local conductivity, $\epsilon $ is the dielectric
constant of SiO$_{2\text{ }}$ and $\phi _{0}(x,t)=\frac{1}{\pi }%
V_{ac}\arctan (x/d)\cos (\omega t)$ \cite{davies}.

\begin{figure}[tbp]
\vbox{\vspace{0 in}} \hbox{\hspace{+0.1in}} \epsfxsize 3.2 in %
\epsfbox{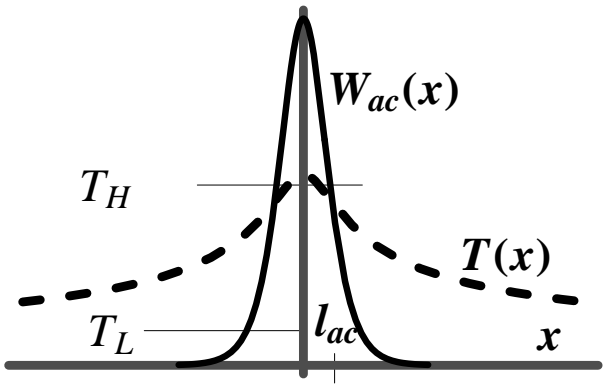 } \vskip 0.5in
\caption{Joule input (solid line) and electron temperature (dashed line) 
\emph{vs} distance $x$ from the center of the hot strip }
\label{tx}
\end{figure}

At high frequencies, $\omega /2\pi >$10 GHz, the major part of the microwave
power is absorbed by the 2DEG in the narrow strip under the slit between the
two gates. The size of the narrow region (hot strip) is 
\begin{equation}
l_{ac}=\left( \frac{2\sigma d}{\omega \epsilon \epsilon _{0}}\right) ^{1/2}.
\label{l_ac}
\end{equation}%
Figure \ref{tx} shows the distribution of microwave power (solid line)
obtained by a numerical solution of Eqs. (\ref{field}-\ref{field3}). The
microwave power is found to be localized near the slit.

The electron temperature distribution obeys the thermoconductivity equation: 
\begin{equation}
\nabla (\kappa \nabla T(x))=F(T(x))-W_{ac}(x)  \label{thermo}
\end{equation}%
where $\kappa $ is thermoconductivity coefficient, $W_{ac}(x)~$is Joule heat
and $F(T)$ stands for the power losses. The result of numerical integration
of Eq. (\ref{thermo}) together with Eqs. (\ref{field}-\ref{field3}), gives a
temperature distribution shown schematically in Fig. \ref{tx} by the dashed
line. We use $F(T)=R(T)-R(T_{L})$ for the power losses due to phonons \cite%
{pudalov}, where $R(T)$ is the electron-phonon relaxation rate and $T_{L}$
is the lattice temperature, which we assume to be unaffected by the
microwaves. Fig. \ref{tx} shows that the Joule heat decreases exponentially
from the slit as $\exp [-2x/l_{ac}]$ , while the electron temperature
relaxes much more slowly as $(x/L_{T})^{-1/2}$.

At zero $dc$ current the temperature gradient and electric field are related
by \cite{abr} 
\begin{equation}
E(x)=Q(T)\nabla T(x),  \label{tpeqn}
\end{equation}%
where the thermopower coefficient $Q=\frac{1}{3}\pi ^{2}\beta T/(eE_{F})$
with numerical coefficient $\beta \sim 1$ \cite{butcher}. The electric
potential difference between the two contacts is found by integrating the
left and right sides of equation (\ref{tpeqn}) over the distance $x$:

\begin{eqnarray}
V_{dc} &=&A(T_{L})\cdot \left( y(0)-1\right)  \label{eqn_rect} \\
A(T_{L}) &=&\frac{\pi ^{2}T_{L}^{2}}{6e}\left( \frac{\beta _{L}}{E_{FL}}-%
\frac{\beta _{R}}{E_{FR}}\right) ,  \label{A}
\end{eqnarray}%
where we have introduced a dimensionless parameter $y(x)=(T(x)/T_{L})^{2} $.
The indexes $L$ and $R$ correspond to the left and right sides of the 2DEG
and $T(x)$ is the temperature of the hot electrons along the sample. The
coefficient $A$ is antisymmetric with respect to the difference between
Fermi energies in the left and right regions and is thus antisymmetric with
respect to the difference of electron densities $\Delta n=n_{R}-n_{L}$. This
behavior is consistent with the experimental observations (see Fig. \ref%
{Rect_n}).

At low $ac$ power a small electron overheating $\Delta T=T(0)-T_{L}$ is
proportional to the microwave power $P$ and the linear dependence of $V_{dc}$
on the microwave power follows from Eq. (\ref{eqn_rect}). In the high power
regime the voltage $V_{dc}$ is determined by the electron temperature below
the slit $T(0)$ which, in turn, depends non-linearly on the microwave power.
The crossover from the linear to the non-linear regime occurs at $%
(T(0)-T_{L})/T_{L}\sim 1$. Note that if $l_{ac}$ is much less than the
sample size, then Eqs. (\ref{eqn_rect}),(\ref{A}) holds independently of the
model for the thermoconductivity and the power loss function $F(T)$.
Therefore, the only parameter that determines the voltage $V_{dc}$ is the
temperature $T(0)$ of the 2D electrons in the hot strip under the slit.

To find the temperature profile we solve the thermoconductivity Eq. (\ref%
{thermo}). Since in our setup the phonon system is not directly heated by
the microwave radiation, the phonon temperature is weakly affected by the
microwaves. To compare theory with the experimental findings, both the
phonon heat transport and the phonon drag contribution to the thermopower
are neglected. To proceed further we use the Wiedemann-Franz law for the
electron thermoconductivity $\kappa (T)=\pi ^{2}\sigma T/(3e^{2})$. When $%
x\gg l_{ac}$ we can reduce the order of the differential equation (\ref%
{thermo}) by neglecting the second term on the right hand side (Joule heat)
yielding 
\begin{equation}
(\frac{\pi ^{2}\sigma T_{L}^{2}}{12\text{\/}e^{2}})\left( \frac{dy}{dx}%
\right) ^{2}=R(T_{L})(\frac{1}{4}y^{4}-y+\frac{3}{4})  \label{y}
\end{equation}%
Experiments \cite{pudalov} have shown that in the temperature range from $1$
to $5$ K 
\begin{equation}
R(T)=R_{1}T^{6},  \label{interpolation}
\end{equation}%
where $R_{1}$ is a constant. For the density $8.5\times 10^{11}$cm$^{-2}$, \ 
$R_{1}\approx 1.3$ mW$\cdot $m$^{-2}$K$^{-6}.~$According to the theory \cite%
{sergeev} $R(T)$ deviates weakly from the $T^{6}$ law and depends on
electron concentration as $n^{-1/2}.$ The temperature relaxation length can
be expressed in terms of $\kappa (T)$ and $R(T)$ as 
\begin{equation}
L_{T}=\sqrt{\frac{T~\kappa (T)}{R(T)}.}  \label{lt}
\end{equation}%
\bigskip

The total Joule heat of the 2DEG $W_{ac}^{T}$ is given by 
\begin{equation}
W_{ac}^{T}=\int \overline{j(x,t)\nabla \phi (x,t)\/dx}=\eta V_{ac}^{2}\left( 
\frac{\sigma _{L}}{l_{ac}^{L}}+\frac{\sigma _{R}}{l_{ac}^{R}}\right) ,
\label{P}
\end{equation}%
where the upper bar denotes average over time and $\eta \sim 1$ is a
numerical factor.

At a stationary state the input heat is partially absorbed by the phonon
system and partially drained via the electron thermal flow. When $l_{ac}\ll
L_{T}$, most of the input power generated in the hot strip, $-l_{ac}<x<l_{ac}
$, must be drained away via the thermal electron flow, because the hot strip
is much smaller than the total overheated area ($\sim L_{T}$) absorbing the
total heat input. In this case, neglecting the phonon absorption in the hot
strip area, a simple intergration of Eq. \ref{thermo} over the region $%
-l_{ac}<x<l_{ac}$ leads to:

\begin{equation}
W_{ac}^{T}=\frac{\pi ^{2}T_{L}^{2}}{6e^{2}}\left[ \sigma _{L}\left( \frac{dy%
}{dx}\right) _{L}-\sigma _{R}\left( \frac{dy}{dx}\right) _{R}\right]
\label{Pflux}
\end{equation}%
Combining (\ref{y}), (\ref{P}), (\ref{Pflux}), (\ref{l_ac}) and (\ref{G}) we
obtain a relation between the microwave input power $P$ and the temperature
of the 2D electrons under the slit as $y(0)=(T(0)/T_{L})^{2}$:

\begin{eqnarray}
P =B(T)\cdot (y(0)^{4}-4y(0)+3)^{1/2}  \label{y0} \\
B(T) =\frac{\sqrt{3}\pi }{6\eta G^2 }\left( \frac{T_{L}^{2}R(T_{L})d}{%
e^{2}\omega \epsilon \epsilon _{0}}\right) ^{1/2}  \label{B}
\end{eqnarray}

These equations, together with equation (\ref{eqn_rect}), determine the
dependence of the rectified voltage $V_{dc}$ on the lattice temperature $%
T_{L}$ and microwave power $P$. Expanding (\ref{y0}) at small $P$ we get

\begin{equation}
y(0)=1+\frac{\eta \sqrt{2}P\,}{\pi G^2 }\left( \frac{T_{L}^{2}R(T_{L})d}{%
e^{2}\omega \epsilon \epsilon _{0}}\right) ^{-1/2}  \label{y0small}
\end{equation}%
and for the weakly nonlinear regime ($y(0)-1\ll 1$) 
\begin{equation}
V_{dc}=\frac{A(T_{L})P}{\sqrt{6}B(T_{L})}\sim \frac{L_{T}}{l_{ac}}\frac{%
eV_{ac}^{2}}{E_{F}}  \label{Vdcsmall}
\end{equation}

One can see that, expressed in normalized values of the DC voltage $%
V_{dc}^{\ast }=V_{dc}/A(T)$ and the input microwave power $P^{\ast }=P/B(T)$%
, the system of equations (\ref{eqn_rect}) and (\ref{y0}) exhibits a
universal form in the whole range of microwave power: 
\begin{eqnarray}
V_{dc}^{\ast } &=&y(0)-1  \label{eqn_rect_norm} \\
P^{\ast } &=&(y(0)^{4}-4y(0)+3)^{1/2}  \label{y0_norm}
\end{eqnarray}%
The universal dependence $V_{dc}^{\ast }$ vs $P^{\ast }$ is plotted in Fig. %
\ref{scaling}, together with the scaled data points. Excellent agreement is
found between the experiment and the theory in a broad range of temperature
and microwave power. Random deviations between the experiment and the theory
observed at low power are mostly due to low signal/noise ratio for signals
below 1 $\mu$V.

To conclude the theory section we consider other mechanisms that may also
lead to rectification of an $ac$ voltage. We first consider the bulk
rectification associated with the fact that $\sigma $ in Eq. (\ref{field})
may vary due to microwave modulation of the electron density and/or mobility
inside the 2DEG \cite{falko,zhang}. To estimate this effect we assume that
the bulk rectification is due to a periodic variation of the electron
density induced by the microwave modulation of the gate voltage $V_{g}$: $%
\delta n=n\cdot V_{ac}/V_{g}$ \cite{zhang}. The induced $dc$ current $%
I_{dc}^{bulk}=\sigma V_{dc}^{bulk}=\delta \sigma V_{ac}$. Therefore $%
V_{dc}^{bulk}=\frac{\delta \sigma }{\sigma }\cdot V_{ac}=\frac{\delta n}{n}%
\cdot V_{ac}=V_{ac}^{2}/V_{g}~$. This is considerably smaller than the
rectification due to the thermoelectric effect: From Eq. (\ref{Vdcsmall}), $%
V_{dc}\sim V_{ac}^{2}/E_{F},$ as $E_{F}\ll V_{g}.$ Comparison of $%
V_{dc}^{bulk}$ with result (\ref{Vdcsmall}) shows that $V_{dc}^{bulk}\sim
2\cdot 10^{-4}V_{dc}$.

Another possible mechanism of rectification is related to the spatial
variation of the electron density across the boundary \cite{zimbov}. Due to
this variation there is a diffusive electron flow through the boundary. The
net flow of electrons must be zero at thermodynamic equilibrium. An internal
electric field $E_{b}$ is established to compensate the diffusive flow
across the boundary creating a so-called contact potential difference.
Microwave radiation moves the electron system away from thermodynamic
equilibrium. The non-equilibrium (symmetric) part of the distribution
function is driven by the internal electric field $E_{b}$ creating a
rectified current and rectified voltage $V_{dc}^{b}$. This voltage is
estimated to be \cite{zimbov}: $V_{dc}^{b}\sim \tau ^{2}e/m\cdot
ln(n_{1}/n_{2})\cdot E_{\omega }^{2}$, where $\tau $ is the relaxation time,
assumed in \cite{zimbov} to be on the same order of\textbf{\ }magnitude as
the transport relaxation time, $m$ is the band mass of the 2D electrons, and 
$E_{\omega }\sim V_{ac}/l_{ac}$ is the $\ ac$ electric field near the
boundary. Using our result for $l_{ac}$ we get%
\begin{equation}
V_{dc}^{b}\sim \frac{1}{en_{1,(2)}d}\tau \omega V_{ac}^{2}\cdot
ln(n_{1}/n_{2}).  \label{zimbovsk}
\end{equation}%
Comparison of $V_{dc}^{b}$ with (\ref{Vdcsmall}) shows that $V_{dc}^{b}\sim
5\cdot 10^{-4}V_{dc}$, and is inconsistent with the significant temperature
dependence observed in the experiment.

\section{Discussion}

In the preceding sections, we reported measurements of the rectification of
microwave radiation at the boundary between two-dimensional electron systems
separated by a narrow gap between independently controlled gates on a
silicon surface. The rectified signal is large, it is odd with respect to
the interchange of the gates and, within a broad range of temperatures and
frequencies, all data for the rectified voltage versus microwave input power
collapse onto a single universal curve using two scaling parameters.

The fact that interchanging the gates gives rise to a signal that is
essentially the same in magnitude and of opposite sign indicates that bulk
rectification of the incoming microwave signal is a small contribution.
Contrary to expectations for bulk rectification, the observed signal changes
sign, does not depend on the detailed geometry of the sample and gates, and
does not depend on the size of the sample. Moreover, as shown in the section
on theory, we estimate that bulk rectification would contribute a signal
that is considerably smaller than the mechanism we propose. It should be
noted further that our observations cannot be attributed to rectification by
non-ohmic contacts, as this would yield a signal that depends on the
microwave field distribution along the sample and, therefore, should not be
simply antisymmetric when the gates are interchanged.

The near-perfect antisymmetry of the rectification on interchanging gates,
the fact that the microwave power is strongly localized near the boundary
between the 2D metals and the excellent agreement with theory, all provide
strong evidence that the observed rectification is an inherent property of
the two adjacent 2D electron systems of different densities. In particular,
we attribute the observed rectification to a thermoelectric response due to
strong local overheating of the electron gas that produces a large thermal
gradient at the gap between the gates, where the electron density changes
abruptly. We now proceed to present a detailed comparison between this
theory and the experimental results.

As shown in Fig. \ref{scaling}, data for the rectified voltage versus input
power can be collapsed over a broad range of temperature and microwave power
onto a single universal curve using two scaling parameters. Plots of Eq. \ref%
{eqn_rect_norm} and Eq. \ref{y0_norm}, are shown by the solid line.
Excellent agreement is obtained between theory and experiment in a broad
range of temperatures and the microwave power.

The scaling coefficients $A(T)$ and $B(T)$ used to obtain the data collapse
vary with temperature, as shown in Fig. \ref{Tdep}. The temperature
dependence of parameter $A(T)$ denoted by the solid line in Fig. \ref{Tdep}
(A), is proportional to $T^{2}$, in agreement with Eq. \ref{A}. At $T=3$ K
the scaling parameter $A(3K)=10\mu $V, in good agreement with the
theoretical estimate obtained using the Fermi energy $E_{F}$ and the
parameter $\beta =$0.2 for $n_{1}=3.67\times 10^{11}cm^{-2}$ for Si-MOSFETs%
\cite{karavolas}. At higher temperatures ($\sim 10$K) the experimental
values of $A$ tend to fall below the theory because of a noticeable part of
the heat emits in the region $\pm l_{ac}$. We also note that for electron
density $n_{1}=3.67\times 10^{11}cm^{-2}$ the Fermi energy in Si-MOSFETS is
about $20$ K so that at $T\sim 10K$ the electron system is not strongly
degenerate and finite temperature corrections to the thermoelectric
coefficient $Q$ (see Eq. \ref{tpeqn}) \ and the Wiedemann-Franz ratio have
to be taken into account.

Surprisingly, the theory gives much better scaling than one would expect
from the accuracy of both the thermopower coefficient $Q$ and the
Wiedemann-Franz relation. This may reflect the fact that acoustic phonon
scattering is quasi-elastic and hot electrons remain in the "energy
quasi-ballistic" regime \cite{Dubrovskii}. A careful analysis of the
electron kinetics is required to verify this.

The scaling parameter $B(T)$ is shown in Fig.\ref{Tdep}(B). The parameter $B$
displays similar behavior as a function of temperature for all measured
frequencies above $4$ GHz. The temperature dependence is due to the strong
dependence of the power losses $F(T)=R_{1}(T^{6}-T_{L}^{6})$ on the lattice
temperature. The solid line shows the theoretical expectation using an
approximation of the power losses by Eq.\ref{interpolation} derived from
recent theory \cite{sergeev} and experiment \cite{pudalov}. At higher
temperatures discrepancies between theory and experiment are seen, which
are, most likely, associated with deviations from scaling at low
frequencies, for reasons discussed below.

Fig.\ref{alpha} shows the temperature dependence of the parameter $\alpha $,
the constant of proportionality that relates the rectified voltage $V_{dc}$
to the microwave power $P$ in the weakly nonlinear regime (see Eq.\ref%
{lin_dep}). In accordance with the theory (see Eq.\ref{Vdcsmall}), the
coefficient $\alpha $ is proportional to the ratio between the scaling
parameter $A$ to the parameter $B$: $\alpha(T)\propto A(T)/B(T)\propto
1/T^{2}$; the theoretically expected behavior is shown in the figure by the
solid straight line. Good agreement with theory is obtained at high
microwave frequencies ($>$10 GHz), where the microwave radiation is well
localized near the boundary between the two 2D metals. However,
progressively stronger deviations from the theory develop as the frequency
is decreased. These deviations correlate with deviations from the scaling
regime observed at frequencies below 4 GHz.

In particular at frequency 0.7 GHz the power dependence of rectification at
high power does not follow the $P^{1/2}$ rule and the universality governed
by Eq. \ref{eqn_rect_norm} and Eq. \ref{y0_norm} is not observed. We suggest
that the observed departures from theoretical expectations are due to the
fact that the experimental results are outside the range of validity of the
theory in its present form. Analytical and numerical estimates indicate that
at low frequency (1 GHz and below) the microwave field is barely localized
near the boundary between the two 2D metals. The corresponding size of the
hot strip at friquency $1GHz$ is $l_{ac}\sim 80$ $\mu $ is considerably
broader than the temperature relaxation length $L_{T}$, especially in the
high temperature domain (several microns at $T>$6 K). For these conditions,
one of the central approximations of the theory ($l_{ac}\ll L_{T}$) is no
longer valid.

\section{Conclusion}

We have measured the rectification of microwave radiation ($0.7$-$20 $ GHz)
at the boundary between two-dimensional electron systems with different
electron densities. For frequencies above $4$ GHz and over a broad range of
temperatures and microwave power, the rectified voltage $V_{dc}$ obeys
two-parameter scaling: the power dependence obtained at different
temperatures and frequencies collapse onto a single universal curve $%
V_{dc}^{*}=f^*(P^{*})$. Over the range investigated in these experiments,
the scaling exhibits two different power regimes. For small power the
voltage is a linear function of power, $V_{dc}^{*} \propto P^{*}$, while at
higher power the rectification is proportional to $(P^*)^{1/2}$. A theory is
proposed that attributes the rectification to the thermoelectric response
caused by strong local overheating of the 2D electrons by the microwave
radiation at the boundary between two dissimilar 2D metals. Excellent
agreement is obtained between theory and experiment.

\begin{acknowledgments}
The work at the City College of New York was supported by DOE grant
DOE-FG02-84-ER45153. The work at International Center of Condensed Matter
Physics, Bras\'{\i}lia, was supported by IBEM fund from Brazilian Ministry
of Science and Technology.
\end{acknowledgments}



\begin{references}
\bibitem{contacts} The data reported in subsequent figures of this paper were obtained for a sample in which the gates, electrical contacts and microwave electric field are aligned as shown in Fig. \ref{fig1_rect}, so that the DC and microwave electric fields are parallel.  Similar results were obtained for  $\Gamma$-shaped gate configurations where the DC electric field changes direction between the contacts.
\bibitem{bfield}  For temperatures $T<$4K the data were obtained in a small magnetic field of $0.4$ T to suppress superconductivity of the aluminum gates and avoid the change (about 20\%- 30\%) of the microwave field $E_\omega$ near the sample due to a superconducting transition \cite{hoxha}.
\end{references}
\end{document}